\documentclass[a4paper,fleqn]{cas-sc}
\usepackage[english]{babel}
\usepackage[utf8]{inputenc}
\usepackage[T1]{fontenc}
\usepackage{amsmath}
\usepackage{graphicx}
\usepackage{tabularx}
\usepackage{float}

\usepackage[numbers]{natbib}
\usepackage[normalem]{ulem}
\usepackage{soul}
\sethlcolor{yellow} 

\usepackage[dvipsnames,svgnames,table]{xcolor}

\usepackage{color}

\shorttitle{The Regressive Truth}
\shortauthors{Dias et al.}

\begin{document}
\title[mode=title]{The Inequity of Consumption-Based Tax Systems} 



\date{\today}
\author[1]{Thiago Dias}[type=author, auid=000, bioid=1,orcid=0000-0003-0641-7686]
\ead{thiagodias@utfpr.edu.br}
\credit{Conceptualization, Software, Methodology, Visualization, Writing---original draft, Writing---review \& editing}
\author[2,3]{José Roberto Iglesias}[type=author, auid=000, bioid=2,orcid=0000-0002-0232-299X]
\credit{Conceptualization, Methodology, Writing---original draft, Writing---review \& editing}
\ead{iglesias@if.ufrgs.br}
\author[2]{Sebastian Gonçalves}[type=author, auid=000, bioid=3,orcid=0000-0002-3100-9126]
\credit{Conceptualization, Methodology, Writing---original draft, Writing---review \& editing}
\ead{sgonc@if.ufrgs.br}

\affiliation[1]{organization = {Universidade Tecnológica Federal do Paraná, Campus Dois Vizinhos}, addressline={Estrada para Boa Esperança, km 04}, postcode ={85660-000}, city={Dois Vizinhos}, state={PR}, country={Brazil}}
\affiliation[2]{organization = {Instituto de Física, Universidade Federal do Rio Grande do Sul}, postcode ={Caixa Postal 15501}, city={Porto Alegre}, state={RS}, country={Brazil}}
\affiliation[3]{organization = {Instituto Nacional de Ciência e Tecnologia de Sistemas Complexos, INCT-SC, CBPF}, city={Rio de Janeiro}, state={RJ}, country={Brazil}}
\date{\today}
\cortext[ca]{Corresponding author}

\begin{abstract}
This study examines the lack of redistributive effectiveness of consumption-based tax systems with respect to social fairness.
Through numerical simulations, we explore the wealth exchanges among economic agents subject to flat consumption taxes, comparing universal redistribution with optimal targeted approaches. The results demonstrate that consumption taxes exhibit inherent regressivity, disproportionately burdening the poorest 40\% of households who contribute over half of total tax revenue for most tax rates.
The findings challenge the equity of consumption taxes and provide quantitative insights for designing more fair fiscal policies.
\end{abstract}


\maketitle

\section{Introduction}
Wealth inequality has been a subject of global concern since the nineteenth century, with early analyses by Marx and Pareto~\cite{capital-marx,pareto-1896}. A certain degree of equity in wealth distribution has been shown to be necessary for maintaining a functional and stable economy~\cite{cardoso-pa-2020}. Nevertheless, disparities between the assets held by the rich and the poor have been increasing worldwide~\cite{credit-suisse22}. Cross-country comparisons of wealth (or income) distributions reveal two broad classes: a wealthy elite, comprising about 5–10\% of the population, whose wealth follows a Pareto power-law distribution; and the remaining majority, whose income is better described by Gibbs or log-normal distributions~\cite{alvaredo2018,saez2016wealth}.

Since asset transactions drive healthy economies, the concentration of wealth within a small elite leads to stagnation and, ultimately, the collapse of an economy. In the twenty-first century, almost all countries have experienced (and continue to experience) increasing concentration of resources~\cite{capitalxxi,oecd-dataexplorer}. In an ideal \textit{laissez-faire} market, even without explicit bias, the rich would systematically accumulate more wealth in the form of money, capital, and property. Analytical work further shows that, under such conditions, all available resources ultimately  concentrate in a single individual~\cite{cardoso-csf-2023}.

Taxation and redistribution are widely employed to curb rising economic inequality. These mechanisms represent a transfer from richer to poorer agents through taxation, helping to prevent economic stagnation. In a previous contribution, we examined the impact of wealth---and consumption---tax systems on inequality~\cite{dias-pa-2024}. We found that, for each tax rate, there exists a redistributional share of the population that minimizes inequality. Moreover, our results showed that wealth-based systems are generally more effective at reducing inequality than consumption-based ones. Nevertheless, most countries adopt the latter~\cite{bunn-ff-2021}, typically directing redistribution to about 10\% of the vulnerable population~\cite{abramovsky-ifs-2017}. 
A recent example comes from France, where the government changed five times in less than two years~\cite{bbc-france}, struggling to pass a budget that reduced workers' incomes and benefits, while resisting the adoption of even a modest tax on the hyper-rich~\cite{lemonde}. This resulted in continuous social unrest.

Consumption taxes, such as the value-added tax (VAT), are often considered unfair because they disproportionately burden lower-income households~\cite{warren-oecd-2008,blasco-jpe-2023}. To mitigate this regressivity, some essential goods primarily consumed by these households are exempted from taxation. 
Nevertheless, some authors defend consumption tax base.
For example, Bradford and Todler~\cite{bradford-nta-1976} argued that this taxation is \textit{neutral between current and future consumption}, since individuals are taxed only when they spend. In practice, this means that the after-tax purchasing power remains the same whether income is consumed immediately or first invested and then consumed after generating returns.
Bankman and Weisbach further compared ideal taxes (\emph{i.e.}, a time-invariant flat tax rate) on consumption and income concluding that the former is preferable because it appropriately charge most forms of wealth input~\cite{bankman-slr-2006}.

Here, we employ a Statistical Physics framework to study wealth transactions at the microscopic level and their impact on the macroeconomic measures. This approach has been shown to successfully capture the complexity of economic data~\cite{moukarzel-epj-2007,chakraborti-qf-2011,bertotti-ejp-2016,cardoso-csf-2023}. In particular, this work analyzes the fairness of a consumption-based tax system. The paper is organized as follows: Section~\ref{sec:model} presents the model; Section~\ref{sec:measures} discusses the motivation for using the Palma index alongside the Gini coefficient; and Sections~\ref{sec:results} and~\ref{sec:conclusions} present the results and conclusions, respectively.

\section{Kinect exchange model with consumption taxes}\label{sec:model}
The model studied here is a variant of the Yard-Sale model, where economic agents engage in wealth exchanges through pairwise interactions. A population of $N = 10^4$ agents is considered, each characterized by two attributes: their wealth $w_i$ and their risk propensity $\alpha_i$. At the beginning of the simulation, both $w_i$ and $\alpha_i$ are randomly assigned. While the risk propensity remains fixed during the dynamics, wealth flows between agents, subject to the conservation rule $\sum_i w_i = 1$. To avoid numerical underflow, a minimum wealth threshold of $w_{\min} = 10^{-12}$ is imposed: if an agent’s wealth falls below this value, it is set to zero, and the agent ceases to participate in further exchanges.

During each Monte Carlo step (MCS), all agents with $w > w_{min}$ are randomly paired, ensuring that every agent engages in exactly one exchange.

Taking agents $i$ and $j$ to engage in an exchange and assuming that agent $i$ wins, the post-transaction wealths are given by
\begin{align}
    w_i(t+1) &= w_i(t) + (1 - \lambda)\,\Delta w, \nonumber\\
    w_j(t+1) &= w_j(t) - \Delta w, \label{exchange}
\end{align}
where $w_{i(j)}(t+1)$ and $w_{i(j)}(t)$ denote the wealth of agent $i$ ($j$) after and before the transaction, respectively, and $\lambda$ is the tax rate. Although it may appear from Eq.~\eqref{exchange} that only the winner is directly taxed, this is not the case: in practice, sellers and workers include the tax in the price of goods and services.

To mimic realistic wealth exchanges, the amount at stake must be the same for both individuals. This is ensured by~\cite{hayes,cardoso-pa-2020}
\begin{equation}
\Delta w = \min\big[\alpha_i w_i(t), \alpha_j w_j(t)\big],
\end{equation}
where $\alpha_{i(j)}$ represents the fraction of wealth each agent risks in the transaction.

After the trade and collecting the taxes, the total fiscal revenue ($\Lambda$) is equally distributed among the agents that belong to the fraction $\tau$ (also referred to as the target) of the population with wealth sorted increasingly. Unlike the trivial case of wealth tax, where the revenue from taxation for a given $\lambda$ during the entire simulation is the same, regardless of $\tau$; in the case of a consumption tax, $\Lambda$ depends on the amount of wealth exchanged at each step. Please see Subsection~\ref{sec:dynamics} for a more detailed discussion of the case of low $\lambda$, where the number of exchanges is severely reduced due to the high inequality.

The experience in some countries shows that high tax rates can trigger evasion or push taxpayers to relocate their businesses to more flexible jurisdictions~\cite{abramovsky-ifs-2017}. In our model, evasion is not possible, as all trades are taxed, and migration is also excluded because the fixed-$N$ constraint prevents agents from moving to other economies.
Moreover, no information about an individual’s wealth is considered when purchasing goods or services, so the tax burden is identical for rich and poor alike. In this work, we apply a flat tax rate with no distinctions by product or service, resembling practices reported in countries such as Estonia, Hong Kong, and Russia~\cite{grecu-2004,keen-imf-2006}.

\section{The Palma Index}\label{sec:measures}

Measuring inequality is fundamental to understand social structures and the dynamics of economic interactions in socio-economic systems. Both traditional economics and econophysics employ diverse methodologies to quantify wealth inequality. Among the various indicators, the Gini coefficient is widely used; it ranges from 0 (perfect equality) to 1 (maximum concentration, where all wealth is held by a single individual). The Gini coefficient, in turn, is derived from the Lorenz curve $L(p)$, which represents the cumulative share of wealth held by the bottom fraction $p$ of the population.

The mathematical expression of the Lorenz curve is~\cite{fellman-tel-2018}:
\begin{equation}
    L(p) = \frac{1}{\langle w\rangle} \int_0^{w_p}wf(w)dw,
    \label{lorenz}
\end{equation}
where $f(w)$ is the wealth distribution function, and $w_p$ satisfies $F(w_p) = p$, with $F(w)$ being the cumulative wealth function. The Gini coefficient $G$ is defined as twice the area between the equality line---a straight line from $(0,0)$ to $(1,1)$---and the Lorenz curve:
\begin{equation}
    G = 1 - 2\int_0^1 L(p) dp.
    \label{gini}
\end{equation}

The Gini coefficient considers the share of wealth of all individuals. However, as Palma noted in his analysis of the income distribution of 112 countries, the middle and upper-middle classes (deciles five to nine) display remarkable homogeneity, consistently holding about 45–55\% of total income~\cite{palma-desa-2006}. This suggests that inequality is primarily driven by the extremes of the distribution: the poorest 40\% and the richest 10\%. To capture these disparities, the Palma index $P$ is used in addition to the Gini coefficient. It is defined as

\begin{equation}
    P = \frac{1 - L(0.9)}{L(0.4)},
    \label{palma}
\end{equation}
where $L(p)$ is the Lorenz curve. The index has no finite upper bound, with $P \to \infty$ as $G \to 1$. In practice, real economies typically exhibit values of $P$ between 0.25 and 7~\cite{palma-desa-2006}. In this work, we set a cutoff at $P=15$: values above this threshold are excluded, since they would correspond to a situation in which the top decile holds 15 times the wealth of the bottom four deciles, an extreme imbalance not observed empirically.

Figure~1(a) illustrates the complementarity of $G$ and $P$. It compares two wealth distributions with the same Gini value ($G=0.375$) but different Palma indexes ($P_A=1.467$ and $P_B=1.671$). While $G$ cannot distinguish between them, $P$ reveals that distribution B has a heavier upper tail. This is also clear in the Lorenz curves shown in Fig.~1(b). Distribution A lies further below the equality line up to $p \approx 0.65$, but in the top decile the situation reverses: the richest 10\% hold about 20\% of total wealth in A, versus more than 30\% in B. Differences are less pronounced at the bottom, where the poorest 40\% own roughly 15\% of wealth in A and 20\% in B.

\begin{figure}[!htbp]
    \centering
    \includegraphics[width = 0.75\textwidth]{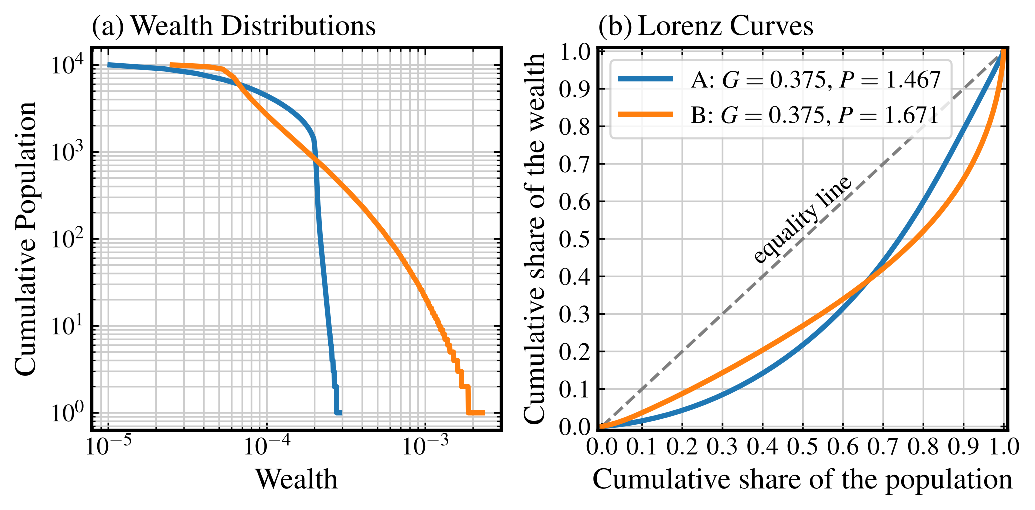}
    \caption{(a) Wealth distributions (log-log scale) for two systems that have the same Gini index $G = 0.375$. (b) Lorenz curves of the same systems. The Palma indices vary more than 10\%, $P_A = 1.467$ and $P_B=1.671$. Simulation parameters: $\lambda=0.79$ and $\tau=0.9$ (A), $\lambda=0.37$ and $\tau=0.22$ (B).}
    \label{fig:dists-lorenz}
\end{figure}

These examples show that the Gini coefficient alone may obscure differences in the tails of the distribution. By focusing on the extremes, the Palma index provides a more sensitive measure of economic imbalance. It also facilitates direct comparisons between the wealth shares of the rich and the poor, which is particularly useful in policy discussions about redistribution and taxation.

\section{Results}\label{sec:results}
\subsection{Dynamics of Low Tax Rates}\label{sec:dynamics}
\begin{figure}[!htbp]
    \centering
    \includegraphics[width = 0.75\textwidth]{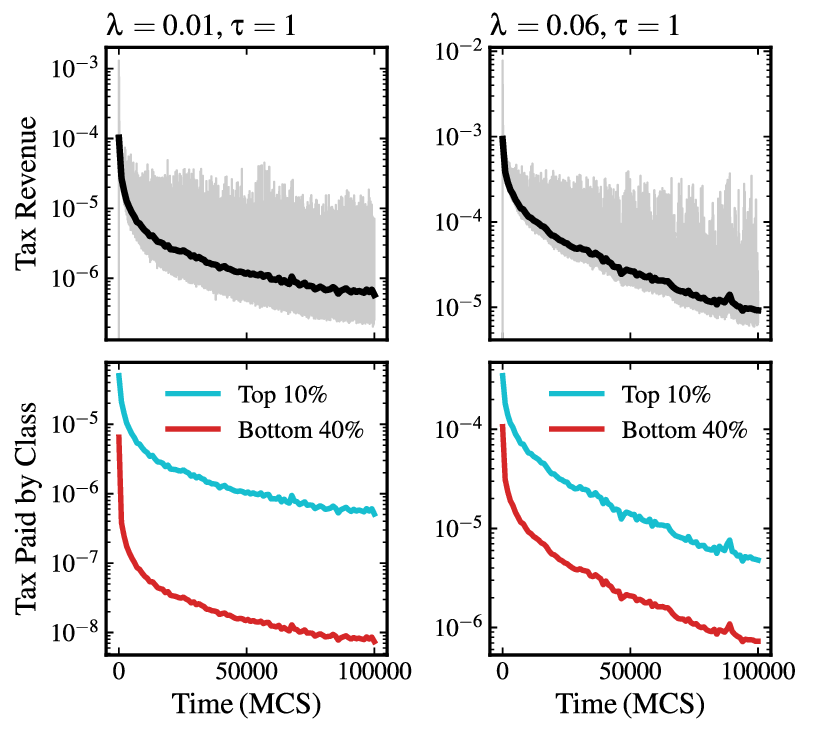}
    \caption{Top: tax revenues ($\Lambda$) as functions of time for two rates, $\lambda = 0.01$ (left) and $0.06$ (right). Bottom: tax paid by each class. The gray lines represent the measures at each MCS, while black, cyan, and red lines indicate averages over $10^3$ steps. Note the different vertical scales.}
    \label{fig:dyn_lowtax}
\end{figure}

We focus here on the low-taxation regime, where highly unequal distributions emerge, as evidenced by the high values of the Palma index ($P > 15$). Therefore, in this case, we resort on the Gini coefficient as the inequality measure instead of the Palma index. 
We present here two measurements (or indices) for two cases: the total tax revenue, $\Lambda$, and the total tax paid by the top 10\% and the bottom 40\% classes. 
Both quantities are expressed in terms of the total wealth of the system---since the total wealth is normalized to one, there is effectively no difference between relative and absolute values.  The two cases correspond to different tax rates applied to exchanges, namely $\lambda = 0.01$ and $\lambda = 0.06$.
Figure~\ref{fig:dyn_lowtax} displays the tax revenue $\Lambda$ over time for two tax rates ($\lambda = 0.01$ and $0.06$) plus universal redistribution ($\tau = 1$), which maximizes $\Lambda$\footnote{See appendices~\ref{sec:contour} and~\ref{sec:table} for details.}. Unlike the wealth-based tax system, where $\Lambda = \lambda$, the exchange-based system yields $\Lambda < \lambda$.

The top panels of Fig.~\ref{fig:dyn_lowtax} show a monotonic decrease in $\Lambda$ for both tax rates. The gray area represents the revenue at each step. The fluctuations arise because, at each MCS, different pairs of agents interact, exchanging varying amounts of wealth and thus generating different revenues.
For example, if rich agents interact among themselves, large amounts are exchanged. By contrast, if rich and poor agents interact, the exchanged amounts are smaller, resulting in lower revenues.  Since the pairs of agents change randomly from step to step, this explains the observed fluctuations. To reduce the impact of them, we perform averages over sets of $10^3$ MCS.

As shown in the plots of the time evolution of $\Lambda$, in both cases it decreases by two orders of magnitude asymptotically. This trend arises from growing inequality: as wealth concentrates within a small group of rich agents, both the frequency and the volume of exchanges decline. For instance, with $\lambda = 0.01$, about 20\% of the agents have zero wealth, while another 20\% (the lowest four deciles) each possess $w_i < 10^{-10}$. In contrast, the richest 10\% hold $w_i > 10^{-6}$. This disparity implies that most transactions take place among low-wealth individuals.

\begin{figure}[!htbp]
    \centering
    \includegraphics[width=0.5\linewidth]{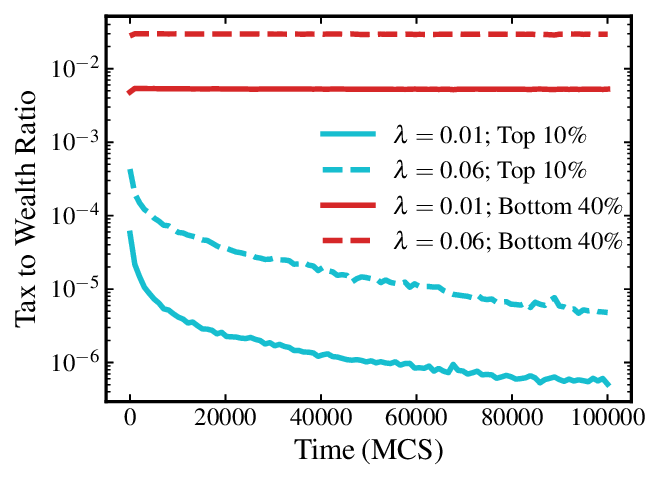}
    \caption{Proportion of wealth paid in taxes for each class. The tax rates are in the legend of the figure and universal redistribution ($\tau = 1$) is considered.}
    \label{fig:dyn_prop}
\end{figure}

The bottom panels of Fig.~\ref{fig:dyn_lowtax} show the evolution of the tax paid by each class. 
At first glance, the system may appear progressive as the richer agents pay roughly 8 times more tax for $\lambda = 0.06$. However, a closer inspection, presented in Fig.~\ref{fig:dyn_prop}, tells a different story. This figure displays the ratio of taxes paid by each class relative to the wealth they hold. For the lower class, about 1\% of their wealth is paid as taxes for $\lambda = 0.01$ and $\lambda = 0.06$. This calculation includes zero-wealth agents, a detail that differs from empirical studies~\cite{blasco-jpe-2023}. In contrast, the upper class pays less than $10^{-5}$ of their wealth in taxes. As wealth accumulates, the richest agents contribute an increasingly smaller fraction of their wealth to taxation, highlighting the system's effective regressivity despite the apparent progressivity in absolute terms.

\begin{figure}[!htbp]
    \centering
    \includegraphics[width=0.5\linewidth]{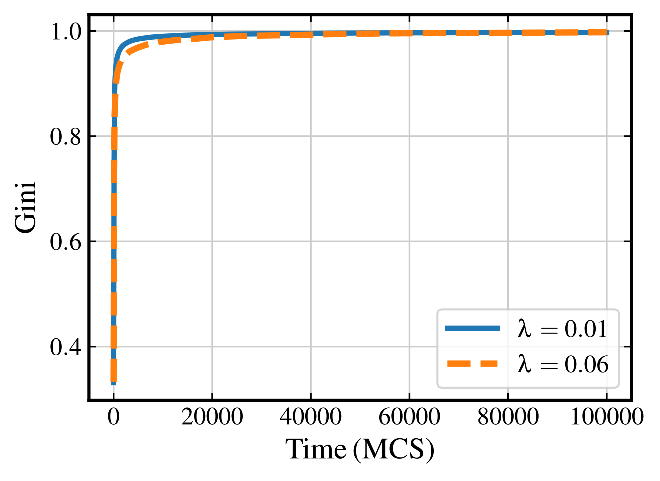}
    \caption{Temporal evolution of the Gini coefficient under low-tax regimes ($\lambda = 0.01$ and $\lambda = 0.06$).}
    \label{fig:dyn_gini}
\end{figure}

One may argue that the system has not yet reached equilibrium, since $\Lambda$ continues to decrease and has not fully stabilized. However, the tax-to-wealth ratio still evolving (decreasing indeed) corresponds to the top 10\% of the population, while the bottom 40\% stabilizes very quickly. Moreover, other measures, such as the Gini coefficient ($G$) and the Palma index ($P$), remain essentially constant. For instance, Fig.~\ref{fig:dyn_gini} shows the temporal evolution of the Gini coefficient in these cases. Given that small tax rates tend to increase inequality slowly over time, all simulations were truncated at $t = 10^5$\,MCS.

\subsection{Inequality and Tax Revenue at Steady State}\label{sec:equilibrium}

When the system reaches a steady state (for tax rates $\lambda \geq 0.22$), the wealth distribution stabilizes, as do the tax-related measures, including total revenue and the tax shares paid by each socioeconomic class.
For regimes with lower tax rates ($\lambda < 0.22)$, the outcomes exhibit significant fluctuations due to the high concentration of wealth, therefore, the following analysis on the tax-related measures focuses on the average values over the last $10^3$ Monte Carlo steps.

\begin{figure}[!htbp]
    \centering
    \includegraphics[width = 0.75\textwidth]{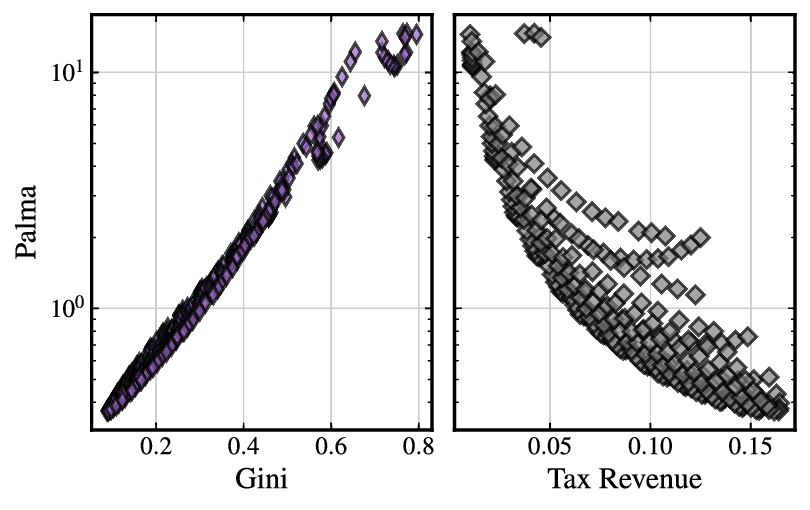}
    \caption{Relationship between inequality measures and taxation outcomes. (Left) Scaling behavior between Palma and Gini indices, showing superlinear growth ($P \sim G^\alpha$). (Right) Dependence of Palma index on tax revenue ($\Lambda$).}
    \label{fig:eq_palma}
\end{figure}

Figure~\ref{fig:eq_palma} presents scatter plots characterizing the relationship between inequality measures and taxation outcomes. The left panel reveals a scaling relationship between the Palma and Gini indices, well-described by a power-law dependence ($P \sim G^\alpha$) in the intermediate Gini range. This superlinear scaling indicates that the Palma index grows disproportionately faster than the Gini coefficient as inequality increases, highlighting its enhanced sensitivity to extreme wealth concentrations.

The right panel of Fig.~\ref{fig:eq_palma} illustrates the connection between tax revenue and inequality reduction. Although the relationship between Palma index and tax revenue ($\Lambda$) is not straightforward, a general trend emerges: as tax revenue increases and redistribution mechanisms take effect, the Palma index decreases, indicating reduced inequality.

\begin{figure}[!htbp]
    \centering
    \includegraphics[width = 0.75\textwidth]{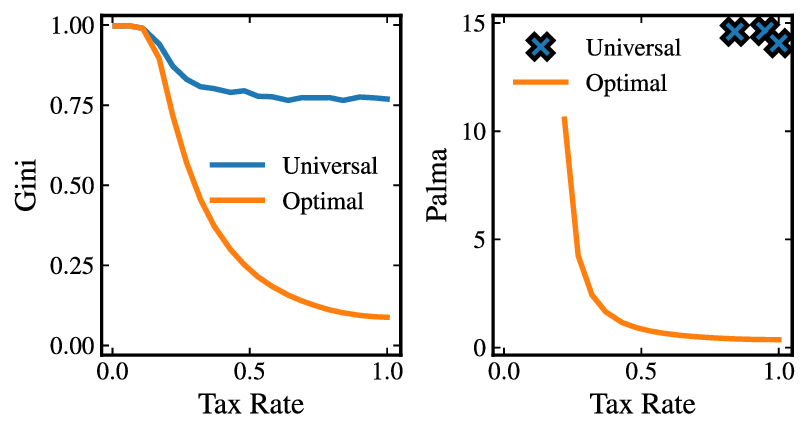}    
    \caption{Comparison of inequality reduction under different redistribution schemes. (Left) Gini coefficient and (right) Palma index as functions of tax rate $\lambda$ for universal redistribution ($\tau = 1$) and optimal targeted redistribution to the poorest $\tau N$ agents.}
    \label{fig:eq_palma-gini}
\end{figure}

Our findings align with previous studies~\cite{dias-pa-2024,iglesias-cnsns-2020,nener-ptrsa-2022} regarding the existence of an optimal redistribution fraction $\tau$ that minimizes inequality. Figure~\ref{fig:eq_palma-gini} compares the behavior of Gini and Palma indices under two redistribution schemes: universal redistribution ($\tau = 1$) and optimal targeted redistribution to the poorest $\tau N$ agents. The results demonstrate that while universal redistribution produces only modest inequality reduction, targeted redistribution achieves pronounced, monotonic decreases in both inequality measures, approaching near-zero values for high tax rates.

Furthermore, our analysis shows that the optimal redistribution fraction $\tau$ is not universal but depends on the metric of interest. The value that minimizes $G$ differs from the one that minimizes the Palma index, which in turn is distinct from the value that maximizes tax revenue ($\Lambda$). This absence of a single optimum highlights a key implication: the definition of ``optimal'' redistribution is inherently normative, depending on whether the policy goal is to reduce overall inequality (Gini), narrow the gap between rich and poor (Palma), or maximize fiscal revenue ($\Lambda$). The contour plots in Appendix~\ref{sec:contour} illustrate the sensitivity of each metric to the redistribution target $\tau$ across different tax rates $\lambda$, highlighting the critical role of $\tau$ in achieving each objective.

\begin{figure}[!htbp]
    \centering
    \includegraphics[width=0.75\textwidth]{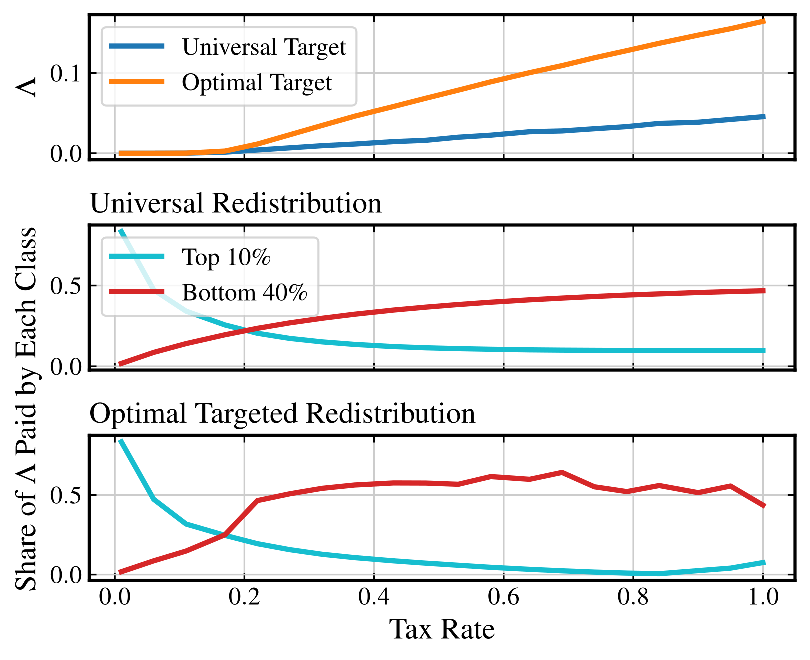}
    \caption{Tax revenue analysis and distribution of tax burden. Top: Tax revenue ($\Lambda$) as function of tax rate ($\lambda$) showing monotonic increase without evidence of Laffer curve behavior. Bottom: Tax burden share by economic class, revealing the regressive nature of consumption taxes.}
    \label{fig:eq_lambda}
\end{figure}

The analysis of tax revenue reveals no evidence of the Laffer curve phenomenon in this consumption-based tax system (Fig.~\ref{fig:eq_lambda}, top panel). Contrary to theoretical predictions~\cite{rondinel-nber-2025}, the tax revenue $\Lambda$ increases monotonically with the tax rate $\lambda$ without exhibiting the characteristic revenue-maximizing turnover point.

Figure~\ref{fig:eq_lambda} (lower panels) reinforces the regressive nature of consumption-based taxation. For tax rates exceeding 0.20 (universal redistribution) and 0.17 (optimal redistribution), the bottom 40\% of households bear the majority of the tax burden. This regressivity is particularly pronounced under optimal redistribution, where the poorest deciles contribute more than half of the total tax revenue for nearly all $\lambda > 0.17$, while the share paid by the top 10\% decreases monotonically with increasing tax rates.

\section{Conclusions}\label{sec:conclusions}
This work has demonstrate that consumption-based tax systems structurally sustain  economic inequality through their regressive nature. 
The computational experiments reveal that taxing wealth exchanges systematically transfers a disproportionate fiscal burden to the lower class. 
The finding that the bottom 40\% contribute to the majority of the tax revenue across virtually all tax rates contrasts with the equity principles that taxation and redistribution are meant to promote.

Economic inequality was analyzed using both the Gini coefficient and the Palma index. The comparison highlights the superiority of the Palma index in capturing small variations in the extreme tails of the wealth distribution. Its use also strengthens the evidence of the regressive behavior inherent in consumption-based taxation.

We also examined the relationship between tax revenue and inequality. In general, as the total revenue $\Lambda$ increases and redistribution takes effect, the Palma index decreases. For low taxation rates, the upper class accounts for the largest share of revenue. Yet, when comparing the levies paid by each class relative to their wealth share, the bottom 40\% bears a tax burden equivalent to about 0.5\% of their wealth at $\lambda = 0.01$ and 3\% at $\lambda = 0.06$, whereas the top 10\% pays less than $10^{-6}$ and $10^{-5}$ of their wealth, respectively.

At equilibrium, even under optimal targeted redistribution, only for $\lambda > 0.22$ do we find $P < 15$. This indicates that consumption taxes are inefficient at preventing the accumulation of wealth among the top 10\% at the expense of the lower classes. For instance, values $P < 2$ are observed only for $\lambda > 0.37$, depending on the redistribution target. These results show that significantly higher tax rates are required to achieve meaningful reductions in inequality.

The sentence above may appear contradictory to which we have stated from the abstract to the first part of our conclusions. The apparent paradox arises because, on the one hand, the results confirm the regressivity of consumption taxes---they impose a heavier relative burden on poorer households and thus challenge the equity of the tax system. On the other hand, the simulations show that reducing inequality requires raising these same taxes substantially (above roughly 40\%), seemingly contradicting the first finding. The paradox resolves when we recognize that such inequality reduction only occurs under full redistribution of the collected tax revenue to the poorest households. In other words, consumption taxes alone are regressive, but when paired with perfectly targeted transfers, they can reduce inequality ---albeit only beyond a high taxation threshold. However, this idealized scenario is rarely observed in real economies where tax revenues are allocated across multiple public sectors beyond redistribution.

Finally, the absence of Laffer curve behavior in consumption-based taxation suggests that tax revenue increases monotonically with the tax rate. This trend becomes more pronounced for $\lambda > 0.17$, where an almost linear relation emerges. However, as the tax rate rises, the share of $\Lambda$ contributed by the upper class decreases in both redistribution schemes, implying that most of the additional revenue originates from the lower class. This exposes the inherent unfairness of consumption taxes---an injustice often overlooked in analyses that focus solely on inequality metrics or aggregate revenue.

Our analysis is limited in that it is based on a simplified economic model---the Yard-Sale model---which assumes pairwise exchanges between identical agents and considers only a consumption-based tax system, applied on transactions. Other forms of taxation, such as those on income, capital, and inheritance~\cite{piketty-nber-2016,bachas-prwp-2020,cantante-hssc-2020,dianov-sus-2022,guvenen-qje-2023,dammerer-cje-2023, revenues-2020}, were not explored here. Extending the present framework to include these alternative systems would provide a broader understanding of how different fiscal mechanisms affect wealth distribution and economic inequality.

\printcredits

\section*{Acknowledgements}
S.G acknowledges support from Conselho Nacional de Desenvolvimento Científico e Tecnológico (CNPq-Brasil) under grant \#  309560/2025-0.
\bibliographystyle{unsrt}
\bibliography{references}

\appendix
\section{Contour Plots of Steady-State Metrics}\label{sec:contour}

\begin{figure}[!htbp]
    \centering
    \includegraphics[width=0.75\textwidth]{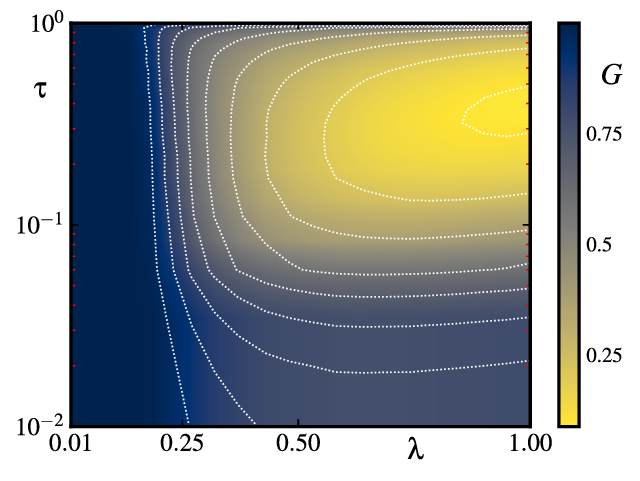}
    \caption{Contour plot of equilibrium Gini as functions of the tax rate $\lambda$ and the fraction $\tau$ of the most vulnerable agents.}
    \label{fig:g_space}
\end{figure}

\begin{figure}[!htbp]
    \centering
    \includegraphics[width=0.75\textwidth]{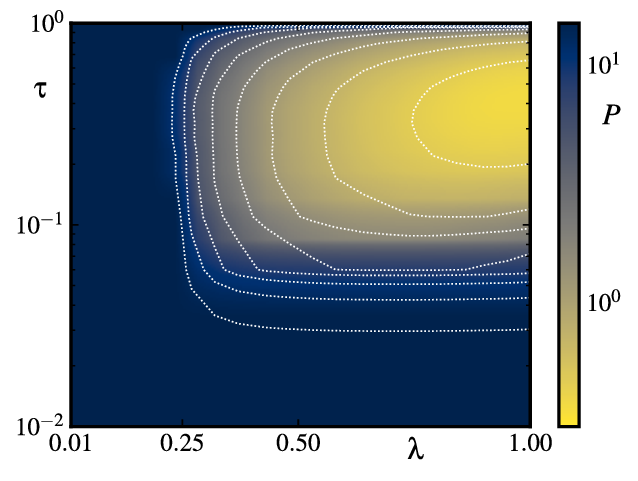}
    \caption{Contour plot of equilibrium Palma as functions of the tax rate $\lambda$ and the fraction $\tau$ of the most vulnerable agents. The colormap is logarithmic, and values of $P>15$ were normalized to $P=15$ for better visualization.}
    \label{fig:p_space}
\end{figure}

\begin{figure}[!htbp]
    \centering
    \includegraphics[width=0.75\textwidth]{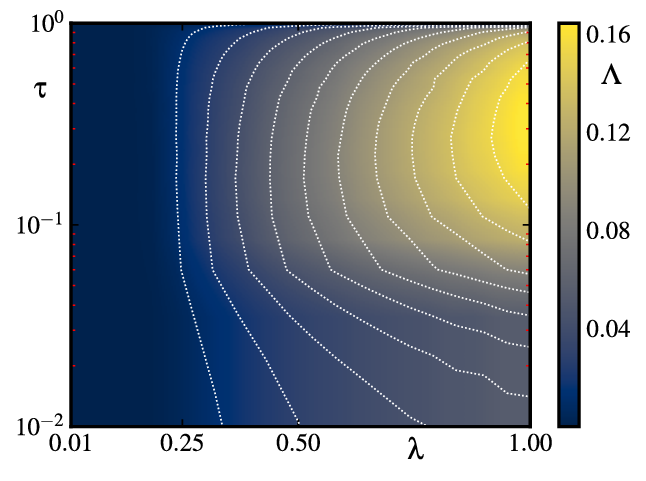}
    \caption{Contour plot of equilibrium tax revenue as functions of the tax rate $\lambda$ and the fraction $\tau$ of the most vulnerable agents.}
    \label{fig:L_space}
\end{figure}

As established in Section~\ref{sec:equilibrium}, the optimal redistribution target is metric-dependent. For completeness, Fig.~\ref{fig:g_space}, Fig.~\ref{fig:p_space}, and Fig.~\ref{fig:L_space} present contour plots for the Gini coefficient ($G$), Palma index ($P$), and total tax revenue ($\Lambda$), respectively. These plots demonstrate that a consumption tax system is inefficient at reducing inequality and increasing total tax revenue when $\lambda < 0.25$, regardless of the redistributional target.

The figures reveal two key features: (i) for tax rates exceeding 0.22, an approximately horizontal path in the $\lambda$-$\tau$ space indicates a nearly unique optimal target for each metric; and (ii) this target fluctuates around 0.32 for both the Gini ($G$) and Palma ($P$) indices, highlighting their scaling relationship. In contrast, for the tax revenue ($\Lambda$), the optimal target remains close to 0.22 for all $\lambda \geq 0.27$.

\section{Optimal targets for each measure of interest}\label{sec:table}

Table~\ref{tab:optimal_vals} reports the optimal target fractions ($\tau$) for the redistribution parameter that respectively minimize the Gini coefficient ($G$), minimize the Palma index ($P$), and maximize the tax revenue ($\Lambda$), for each tax rate ($\lambda$) studied. The table also presents the corresponding achieved values for the three metrics.

\begin{table}[!b]
    \centering
    \begin{tabularx}{\textwidth}{@{}XX@{}}
    \begin{tabular}{c|l|c|c}
    \toprule
        Tax Rate & Measure & Opt. target & Opt. Value \\\midrule
         & $G$ & 1.0 & 0.997 \\\cline{2-4}
        0.01 & $P$ & 0.58 & $-$ \\\cline{2-4}
         & $\Lambda$ & 1.0 & $1.0\cdot10^{-6}$ \\\midrule
         & $G$ & 0.74 & 0.998 \\\cline{2-4}
        0.06 & $P$ & 0.74 & $-$ \\\cline{2-4}
         & $\Lambda$ & 1.0 & $9.0\cdot10^{-6}$ \\\midrule
         & $G$ & 0.95 & 0.990 \\\cline{2-4}
        0.11 & $P$ & 0.43 & $-$ \\\cline{2-4}
         & $\Lambda$ & 0.95 & $1.4\cdot10^{-4}$ \\\midrule
         & $G$ & 0.95 & 0.896 \\\cline{2-4}
        0.17 & $P$ & 0.48 & $-$ \\\cline{2-4}
         & $\Lambda$ & 0.79 & $2.7\cdot10^{-3}$ \\\midrule
         & $G$ & 0.58 & 0.716 \\\cline{2-4}
        0.22 & $P$ & 0.32 & 10.558 \\\cline{2-4}
         & $\Lambda$ & 0.32 & $1.1\cdot10^{-2}$ \\\midrule
         & $G$ & 0.37 & 0.570 \\\cline{2-4}
        0.27 & $P$ & 0.37 & 4.249 \\\cline{2-4}
         & $\Lambda$ & 0.17 & $2.3\cdot10^{-2}$ \\\midrule
         & $G$ & 0.27 & 0.457 \\\cline{2-4}
        0.32 & $P$ & 0.32 & 2.437 \\\cline{2-4}
         & $\Lambda$ & 0.17 & $3.5\cdot10^{-2}$ \\\midrule
         & $G$ & 0.27 & 0.373 \\\cline{2-4}
        0.37 & $P$ & 0.27 & 1.654 \\\cline{2-4}
         & $\Lambda$ & 0.17 & $4.6\cdot10^{-2}$ \\\midrule
         & $G$ & 0.27 & 0.299 \\\cline{2-4}
        0.43 & $P$ & 0.32 & 1.156 \\\cline{2-4}
         & $\Lambda$ & 0.17 & $5.8\cdot10^{-2}$ \\\midrule
         & $G$ & 0.22 & 0.252 \\\cline{2-4}
        0.48 & $P$ & 0.27 & 0.93 \\ \cline{2-4}    
         & $\Lambda$ & 0.17 & $6.8\cdot10^{-2}$ \\\bottomrule
    \end{tabular}
    &
    \begin{tabular}{c|l|c|c}
    \toprule
        Tax Rate & Measure & Opt. target & Opt. Value \\\midrule
         & $G$ & 0.27 & 0.214 \\\cline{2-4}
        0.53 & $P$ & 0.27 & 0.772 \\\cline{2-4}
         & $\Lambda$ & 0.17 & $7.9\cdot10^{-2}$ \\\midrule
         & $G$ & 0.27 & 0.185 \\\cline{2-4}
        0.58 & $P$ & 0.32 & 0.664 \\\cline{2-4}
         & $\Lambda$ & 0.22 & $8.9\cdot10^{-2}$ \\\midrule
         & $G$ & 0.27 & 0.157 \\\cline{2-4}
        0.64 & $P$ & 0.32 & 0.569 \\\cline{2-4}
         & $\Lambda$ & 0.22 & $1.0\cdot10^{-1}$ \\\midrule
         & $G$ & 0.32 & 0.139 \\\cline{2-4}
        0.69 & $P$ & 0.32 & 0.511 \\\cline{2-4}
         & $\Lambda$ & 0.27 & $1.1\cdot10^{-2}$ \\\midrule
         & $G$ & 0.32 & 0.124 \\\cline{2-4}
        0.74 & $P$ & 0.32 & 0.466 \\\cline{2-4}
         & $\Lambda$ & 0.22 & $1.2\cdot10^{-1}$ \\\midrule
         & $G$ & 0.32 & 0.111 \\\cline{2-4}
        0.79 & $P$ & 0.22 & 0.431 \\\cline{2-4}
         & $\Lambda$ & 0.17 & $1.3\cdot10^{-1}$ \\\midrule
         & $G$ & 0.32 & 0.102 \\\cline{2-4}
        0.84 & $P$ & 0.37 & 0.403 \\\cline{2-4}
         & $\Lambda$ & 0.27 & $1.4\cdot10^{-1}$ \\\midrule
         & $G$ & 0.27 & 0.214 \\\cline{2-4}
        0.90 & $P$ & 0.37 & 0.380 \\\cline{2-4}
         & $\Lambda$ & 0.27 & $1.4\cdot10^{-1}$ \\\midrule
         & $G$ & 0.37 & 0.090 \\\cline{2-4}
        0.95 & $P$ & 0.37 & 0.370 \\\cline{2-4}
         & $\Lambda$ & 0.32 & $1.6\cdot10^{-1}$ \\\midrule
         & $G$ & 0.37 & 0.088 \\\cline{2-4}
        1.0 & $P$ & 0.43 & 0.367 \\\cline{2-4}
         & $\Lambda$ & 0.27 & $1.6\cdot10^{-1}$ \\\bottomrule
    \end{tabular}
    \end{tabularx}
    \caption{Optimal targets for each tax rate that minimize Gini index ($G$) and Palma ratio ($P$), as well as maximize the tax revenue ($\Lambda$)}
    \label{tab:optimal_vals}
\end{table}

\end{document}